\documentclass[manuscript]{acmart}
\usepackage{enumitem}
\usepackage{multirow}
\usepackage{booktabs}
\usepackage{graphicx}
\usepackage{verbatim}
\usepackage{vcell}
\usepackage{multicol}
\usepackage{subcaption} 
\AtBeginDocument{%
  \providecommand\BibTeX{{%
    \normalfont B\kern-0.5em{\scshape i\kern-0.25em b}\kern-0.8em\TeX}}}

\copyrightyear{2023}
\acmYear{2023}
\setcopyright{rightsretained}
\acmConference[RecSys '23]{Seventeenth ACM Conference on Recommender Systems}{September 18--22, 2023}{Singapore, Singapore}
\acmBooktitle{Seventeenth ACM Conference on Recommender Systems (RecSys '23), September 18--22, 2023, Singapore, Singapore}
\acmDOI{10.1145/3604915.3608793}
\acmISBN{979-8-4007-0241-9/23/09}

\begin{document}

\title{Deep Situation-Aware Interaction Network for Click-Through Rate Prediction}


\author{Yimin Lv}
%\authornote{Work done when Yimin Lv was an intern at Meituan.}
\affiliation{
  \institution{Institute of Software, Chinese Academy of Sciences,}
  \institution{University of Chinese Academy of Sciences}
  \city{Beijing}
  \country{China}
  }
\email{lvyimin21@mails.ucas.ac.cn}

\author{Shuli Wang}
\affiliation{
  \institution{Meituan}
  \city{Beijing}
  \country{China}
  }
\email{shuliw1996@gmail.com}

\author{Beihong Jin}
\authornote{Beihong Jin is the corresponding author.}
\affiliation{
  \institution{Institute of Software, Chinese Academy of Sciences,}
  \institution{University of Chinese Academy of Sciences}
  \city{Beijing}
  \country{China}
  }
\email{Beihong@iscas.ac.cn}

\author{Yisong Yu}
\affiliation{
  \institution{Institute of Software, Chinese Academy of Sciences,}
  \institution{University of Chinese Academy of Sciences}
  \city{Beijing}
  \country{China}
  }
\email{yuyisong20@mails.ucas.ac.cn}

\author{Yapeng Zhang}
\affiliation{
  \institution{Meituan}
  \city{Beijing}
  \country{China}
  }
\email{zhangyapeng05@meituan.com}

\author{Jian Dong}
\affiliation{
  \institution{Meituan}
  \city{Beijing}
  \country{China}
  }
\email{dongjian03@meituan.com}

\author{Yongkang Wang}
\affiliation{
  \institution{Meituan}
  \city{Beijing}
  \country{China}
  }
\email{wangyongkang03@meituan.com}

\author{Xingxing Wang}
\affiliation{
  \institution{Meituan}
  \city{Beijing}
  \country{China}
  }
\email{wangxingxing04@meituan.com}

\author{Dong Wang}
\affiliation{
  \institution{Meituan}
  \city{Beijing}
  \country{China}
  }
\email{wangdong07@meituan.com}

\renewcommand{\shortauthors}{Y. Lv and S. Wang, et al.}

\begin{abstract}
User behavior sequence modeling plays a significant role in Click-Through Rate (CTR) prediction on e-commerce platforms. Except for the interacted items, user behaviors contain rich interaction information, such as the behavior type, time, location, etc. However, so far, the information related to user behaviors has not yet been fully exploited. In the paper, we propose the concept of a situation and situational features for distinguishing interaction behaviors and then design a CTR model named Deep Situation-Aware Interaction Network (DSAIN). DSAIN first adopts the reparameterization trick to reduce noise in the original user behavior sequences. Then it learns the embeddings of situational features by feature embedding parameterization and tri-directional correlation fusion. Finally, it obtains the embedding of behavior sequence via heterogeneous situation aggregation. We conduct extensive offline experiments on three real-world datasets. Experimental results demonstrate the superiority of the proposed DSAIN model. More importantly, DSAIN has increased the CTR by 2.70\%, the CPM by 2.62\%, and the GMV by 2.16\% in the online A/B test. Now, DSAIN has been deployed on the Meituan food delivery platform and serves the main traffic of the Meituan takeout app. Our source code is available at https://github.com/W-void/DSAIN. 
\end{abstract}




\received{20 February 2007}
\received[revised]{12 March 2009}
\received[accepted]{5 June 2009}

\maketitle

\section{Introduction}
CTR prediction is a critical task in both recommendation systems and online advertising, and improving the accuracy of the CTR prediction has been receiving attention from both academia and industry.

Recently, modeling user behavior sequences \citeN{zhou2018din, zhou2019dien, feng2019dsin, pi2019mimn, pi2020sim, chen2021eta, li2023decision, li2023context} has been introduced into CTR prediction. Generally, a user behavior sequence refers to a sequence of traces of a user performing actions on the app provided by an online service. For an e-commerce platform, a user behavior sequence at least contains the items that the user interacts with. For a specific scenario, a user behavior sequence can be augmented to include the side information of the interaction. Besides item-related and user-related side information, there are three types of behavior-related side information, i.e., the type of the specific behavior, and the time and (physical or virtual) location the specific behavior occurs. 

Early work on user behaviors (e.g., GRU4Rec \cite{hidasi2015session}, Caser \cite{tang2018personalized}) adopts Recurrent Neural Networks (RNNs) or Convolutional Neural Networks (CNNs) to capture the dependency relationships implied in the user behavior sequences. However, these models cannot obtain the semantics of user behaviors due to limited behavior information (e.g., only item ID is available). Recent work introduces the side information of an interaction behavior to understand the user behaviors. Some work (e.g., \citeN{cui2021st, lin2022spatiotemporal, singer2022trans2D, liu2021noninvasive, zhou2020s3, xie2022dif}) exploits temporal and spatial information of behaviors to help model user interest. Some work (e.g., \citeN{gao2019nmtr, gu2020dmt, wu2022feedrec}) extracts behavior patterns and captures complicated correlations among different behaviors from a multi-behavior sequence. However, existing models focus on modeling a certain aspect of side information (such as behavior type or temporal information). No work can simultaneously incorporate behavior types and behavior spatio-temporal information in modeling user behavior sequences, which leads to the waste of some valuable side-information and neglects complex interdependencies among different side information of behaviors. 

We argue that for an interaction of a user with an item, the behavior type, temporal features, and spatial features make up a situation of the interaction, and they are collectively called situational features. Taking a takeout scenario as an example, temporal features, which depict the temporal context in which the behavior occurred, include the hour of the day, the meal time period of day (e.g., breakfast, lunch, dinner, etc.), and the period of week (e.g., weekday and weekend). Nevertheless, when the geographic locations of users are not available, we can treat an app as a virtual space, thus features of the location where a user triggers an action can be listed under spatial features. In particular, we have one feature to record the location of triggering an action (on the search results page or the recommendation list page) and another feature to indicate whether the location of performing an action is an advertising place. Obviously, each situation has its own semantics. 

For modeling the sequences of user behaviors with situational features, we have to face the following challenge. Given a user behavior and the corresponding situational features, how do we generate the high-quality embeddings for these situational features and the behavior, if taking into account multiple internal correlations, including but not limited to the correlations between different situational features of the same behavior? In this paper, we devise a novel CTR model named DSAIN (Deep Situation-Aware Interaction Network), which deals with the above challenge efficiently and elegantly. Our main contributions can be summarized as follows.
\begin{itemize}[leftmargin=*]
\item We identify the noise in user behavior sequences and utilize the reparameterization trick to reduce noise interference. 
\item We combine commonalities and differences in situational features of the same type into their embeddings, and parameterize the embeddings of situational features to learn the approximated interaction between item ID and each situational feature, obtaining refined embeddings of situational features.
\item We devise a light-weighted scheme of modeling tri-directional correlations to coherently capture cross-behavior, cross-situational feature, and cross-channel correlations, obtaining enhanced embeddings of situational features.
\item We conduct extensive offline experiments on three real-world datasets (i.e., two public and one industrial datasets) and an online A/B test. Offline experimental results demonstrate the proposed DSAIN model achieves state-of-the-art performance. Results from the online A/B test show that DSAIN increases the CTR by 2.70\%, the CPM by 2.62\%, and the GMV by 2.16\%. 
\end{itemize}

The rest of the paper is organized as follows. Section 2 introduces the related work. Section 3 describes the DSAIN model in detail. Section 4 gives the experimental evaluation. Finally, the paper is concluded in Section 5.

\section{Related Work}
Our work is related to the research under two closely related topics: CTR prediction and user behavior modeling.
\subsection{CTR Prediction}

Traditional CTR models learn feature interactions with the aid of deep neural networks. The representative work at this stage includes Wide\&Deep \cite{cheng2016wide}, PNN \cite{qu2016product}, DeepFM \cite{guo2017deepfm}, xDeepFM \cite{lian2018xdeepfm}, ONN \cite{yang2020onn} and CAN \cite{bian2022can}.

Benefiting from the increase in computing power and model capacity, many efforts are put into sequential behavior modeling, which aims to infer the user interest with respect to the candidate based on the user historical behavior sequences. Specifically, DIN \cite{zhou2018din} employs an attention mechanism to dynamically reweight a user's historical behavior w.r.t. the candidate. Following this line, DIEN \cite{zhou2019dien} introduces an interest-evolving layer to capture the interest-evolving process with the combination of attention and GRU.  To further consider the intrinsic session structure of user behaviors, DSIN \cite{feng2019dsin} uses a self-attention and a Bi-LSTM to extract in-session interest representations and cross-session evolution patterns, respectively. Later, Transformer-based models such as BST \cite{chen2019bst} also achieve great success in CTR prediction. 

With the ever-growing volume of user interactions, many researchers pay attention to capturing user interest from long behavior sequences, which enables recommender systems to mine long-term behavior dependencies and the periodicity of user behaviors. Representative work in this field can be roughly divided into two categories, i.e., memory-augmented methods (e.g., HPMN \cite{ren2019lifelong} and MIMN \cite{pi2019mimn})  and user behavior retrieval-based methods (e.g., UBR \cite{qin2020user}, two-phase model SIM \cite{pi2020sim}, end-to-end behavior retrieval model ETA \cite{chen2021eta}, and simple hash sampling-based approach SDIM \cite{cao2022sampling}).

\subsection{User Behavior Modeling}
Originating from the conventional user behavior research on simple user behaviors, researchers have extended to further utilize more diverse and richer user behavior data to improve user behavior modeling performance. Recent research on behavior modeling can be roughly divided into two directions, and then we briefly introduce the main work in these two directions as follows.

The first direction is $\textit{increased multiplicity}$ - oriented user behavior modeling, i.e., multi-behavior recommendation. For instance, NMTR \cite{gao2019nmtr} first separately predicts the user-item interactions concerning different behavior types, then assembles them in a cascading manner to account for the cross-type behavior relations. Based on the concatenation of representations of single-type sequences, DMT \cite{gu2020dmt} devises the deep multifaceted Transformers to simultaneously model users' multiple types of behaviors and utilizes MMoE (Multi-gate Mixture of Experts) to optimize multiple objectives. Taking into account complicated relationships among different behaviors as well as the behavior priority, FeedRec \cite{wu2022feedrec} proposes a strong-to-weak attention network that uses strong feedback to distill accurate positive and negative user interest from weak feedback. Furthermore, researchers attempt to explore the multiplex user-item interactive semantics with graph learning techniques in MBGCN \cite{jin2020mbgcn}, GHCF \cite{chen2021ghcf}, MB-GMN \cite{xia2021mbgmn}.

The second direction is $\textit{growing heterogeneity}$ - oriented user behavior modeling. Most of the above-mentioned behavior modeling methods underuse the rich side information of interaction behaviors. To fill this gap, some work further takes into consideration the heterogeneous and diverse features associated with interactions. For instance, Trans2D \cite{singer2022trans2D} utilizes a modified Transformer with 2D self-attention to capture changes in attributes over time. To avoid fusing side information straightforward, which may deteriorate the original item ID representations, NOVA \cite{liu2021noninvasive} proposes to leverage side information as an auxiliary for the self-attention module to learn a better attention distribution, instead of being fused into item representations. DIF-SR \cite{xie2022dif} argues that integrating side information before the attention will limit the learning of attention matrices, thus decoupling various side information with separate attention calculations to further improve NOVA. Moreover, CARCA \cite{rashed2022carca}, FDSA \cite{zhang2019feature}, S3Rec \cite{zhou2020s3}, and MISS \cite{guo2022miss} also achieve satisfied performance in behavior modeling with heterogeneous features.

Compared to existing work, DSAIN introduces the situation and situational features for each behavior, providing a new perspective on user behaviors. Further, DSAIN gives a simple yet effective method to integrate situational features into behavior embedding.

\section{Methodology}
%In this section, we first describe the proposed DSAIN model in detail. Except for the embedding layer, DSAIN includes four modules, i.e., the Behavior Denoising Module (BDM), the Situational Feature Encoder (SFE), the Correlation Fusion Module (CFM), and the Situation Aggregation Module (SAM). At the end of this section, we present the optimization objective of DSAIN. Figure \ref{model architecture} gives the architecture of DSAIN. 
In this section, we first give the problem formulation and then describe the proposed DSAIN model in detail. At the end of this section, we present the optimization objective and analyze the complexity of DSAIN. 

\begin{figure*}[htbp]
    \centering{\includegraphics[width=0.78\textwidth]{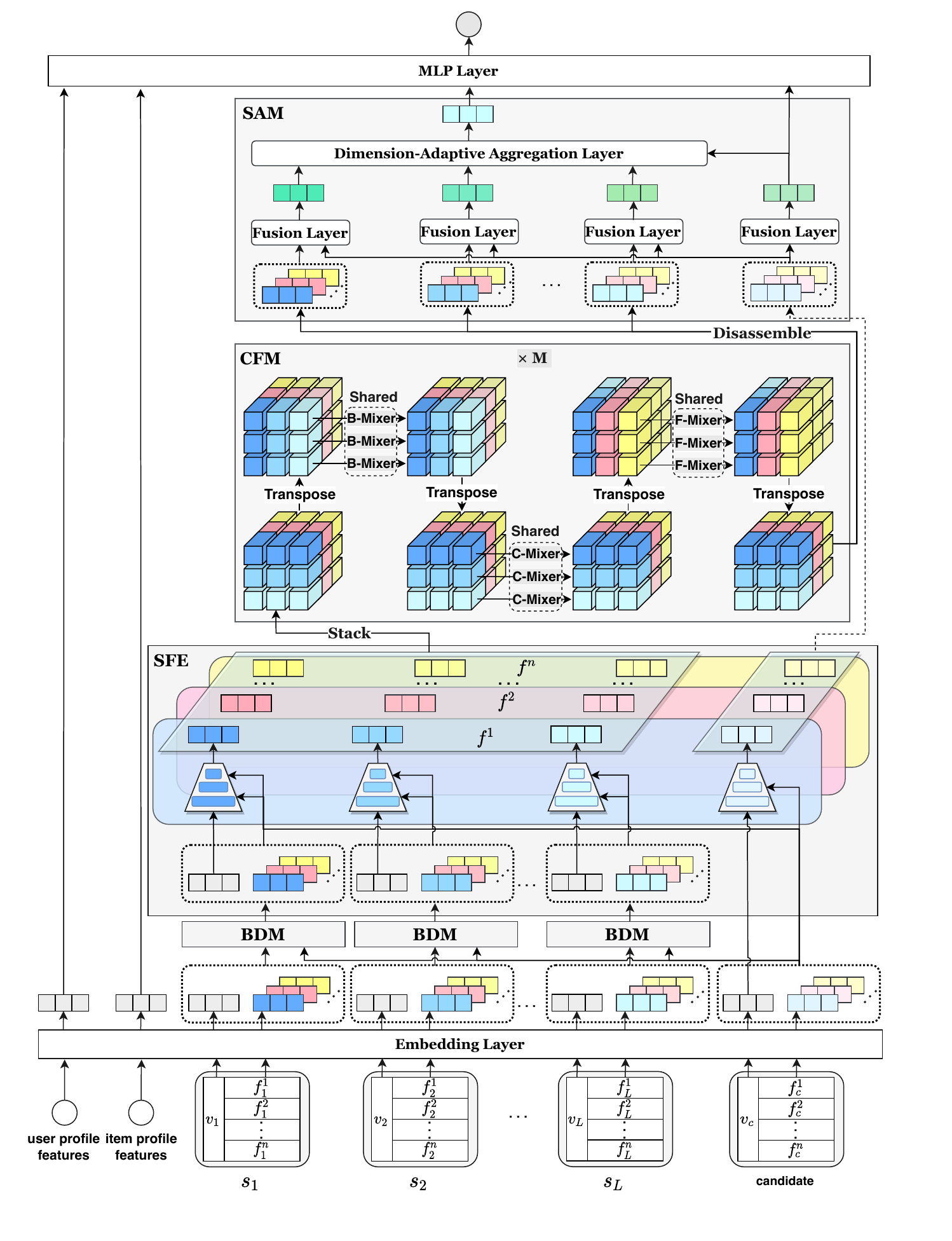}}
    \caption{Architecture of our DSAIN (Deep Situation-Aware Interaction Network). }
    \label{model architecture}
\end{figure*}

Fig. \ref{model architecture} gives the architecture of DSAIN. Except for the embedding layer, DSAIN includes four modules, i.e., the Behavior Denoising Module (BDM), the Situational Feature Encoder (SFE), the Correlation Fusion Module (CFM), and the Situation Aggregation Module (SAM). 

\subsection{Problem Formulation}
% As defined above, behavior dynamic features, including dynamic temporal features, dynamic spatial features, and behavior type, jointly portray the contextual environment in which behavior occurs from multiple perspectives. We define this comprehensive environment as a new concept, named the behavior situation. 

Let $U$, $V$ denote the user set and the item set, respectively. For a user $u \in U$, his/her historical behavior sequence is denoted as $S=\{s_i\}_{i=1}^L$, where $L$ is the length of the behavior sequence, $s_i=(v_i, B_i)$ is the $i$-th behavior, $v_i \in V$ is the interacted item, and $B_i$ is the situation w.r.t. the behavior $s_i$. Supposing we have $n$ types of situational features in total, then the situation $B_i$ can further be represented as $\{f_i^k\}_{k=1}^n$, where $f_i^k$ represents the $k$-th situational feature of the behavior $s_i$. Furthermore, we adopt the context of the current request to represent the current situation for clicking on the candidate $v_c$, denoted as $C=\{f_c^k\}_{k=1}^n$.

Given $\mathcal{X}=\{(u, S)\} \cup \{v_c, C\}$ as input and $y \in \{0,1\}$ as the label of clicking the candidate, our CTR prediction task can be formalized as:
\begin{equation}
\begin{aligned}
\mathcal{P}(y=1|\mathcal{X})= F(\mathcal{X};\theta)
\end{aligned}
\label{CTR prediction task}
\end{equation}	
where $F$ is the probability value output by the network we will develop and $\theta$ represents the parameters of the network. 

\subsection{Embedding Layer}
There exist four groups of features, i.e., user profiles, item profiles, behavior sequences, and the current request context, used as the input of the embedding layer. Specifically, the user profile includes user ID, user age, user gender, etc. The item profile refers to item ID, category ID, etc. The user behavior sequence consists of interacted items and behaviors with corresponding situational features as defined above. Further, the current request context refers to the situation of the target behavior, i.e., the click on the candidate. 
% Concretely, we maintain two different embedding tables with distinct hidden dimensions for situational features and common features, respectively. Transformed by two embedding tables, we can obtain 
Specifically, we utilize two different embedding tables to generate embeddings for situational features and common features, respectively. By looking up embedding tables, the embedding layer outputs the initial embedding vectors including user embedding $\mathbf{u}$, historical interacted items embeddings $\mathbf{v}_i$, behavior situation embeddings $\mathbf{B}_i=\{\mathbf{f}_i^k\}_{k=1}^n$, candidate item embedding $\mathbf{v}_c$, and the current situation embedding $\mathbf{C}=\{\mathbf{f}_c^k\}_{k=1}^n$. Specifically, the dimensions of situational feature embeddings are $d_1$, while others are $d_2$. 

\subsection{Behavior Denoising Module}
Except for clicked items, the user historical behavior sequence considered in this paper is heavily mixed with abundant exposed items, which are imposed on the user by online e-commerce services rather than stemming from the user’s active selection. Although these exposed items can reflect the user interest to a certain extent, there exists much noise which may deteriorate the performance of the CTR model. Therefore, decreasing noise interference in the user behavior sequence by some mechanism is of vital importance.

We start by analyzing a simple filter mechanism. Instead of taking some features such as the category and brand to pick out the top-$k$ items most relevant to the candidate from the user historical behavior sequence, we calculate the correlation between the item and candidate in a soft mode and then sample items based on probabilities. The calculation process of the correlation score is shown in Eq. \ref{keep-prob}.
% Concretely, we first perform an element-wise product between the candidate behavior situation and each item behavior situation and feed it into a single-layer MLP. Then, we adopt the residual connection to emphasize the original information of the item and candidate. Finally, we feed the addition result into another single-layer MLP and then employ the $\operatorname{sigmoid}$ activation function to calculate the correlation score, as shown in Eq. \ref{keep-prob}.

\vspace{-7pt}
\begin{equation}
\begin{aligned}
p_i = \operatorname{sigmoid}\left( \mathbf{W}_2\left(\mathbf{v}_i +  \mathbf{W}_1\left((\mathbf{f}_c^k)\|_{k=1}^n \otimes (\mathbf{f}_i^k)\|_{k=1}^n\right) + \mathbf{v}_c \right)\right)
\end{aligned}
\label{keep-prob}
\end{equation}

\noindent where $\mathbf{W}_1 \in \mathbb{R}^{d_2 \times (n \times d_1)}$ and $\mathbf{W}_2 \in \mathbb{R}^{1 \times d_2}$ are learnable parameters. $\otimes$ means the element-wise product. $(\mathbf{f}_c^k)\|_{k=1}^n$ and $(\mathbf{f}_i^k)\|_{k=1}^n$ concatenate $n$ situational features of the target click and historical behavior $s_i$, respectively. $p_i$ denotes the correlation score of the item $v_i$ w.r.t. the candidate $v_c$, which is adopted as the keeping probability of $v_i$. Then, the probability of discarding the item $v_i$ is $1-p_i$. 
% For brevity, we organize $p_i$ and $1-p_i$ into a list $\delta_i$, i.e., $\delta_i=[p_i, 1-p_i]$. 
Furthermore, we generate a binary variable named selection factor, denoted as $d_i \in \{0,1\}$, $d_i=1$ represents that $v_i$ is reserved and $d_i=0$ means that $v_i$ is discarded.

Then, a feasible way of behavior denoising is sampling from $[p_i, 1-p_i]$ and drawing a conclusion of keeping or dropping the item $v_i$ to generate $d_i$. However, it should be noted that the sampling process is discontinuous and non-differentiable, which is a stumbling block in the model training process based on gradient back-propagation. Inspired by \cite{jang2016categorical}, we leverage the reparameterization trick with Gumbel-Max and then transition to its smooth approximation Gumbel-Softmax, thereby transferring the non-differentiable sampling process into a continuous and differentiable sampling from the Gumbel distribution. The calculation equation is as follows.

\begin{equation}
\begin{aligned}
\hat{d}_i = \frac{\exp \left((g_0 + \log p_i) / \tau \right)}{\exp \left((g_0 + \log p_i) / \tau \right) + \exp \left((g_1 + \log (1-p_i)) / \tau \right)}
\end{aligned}
\label{gumbel-softmax}
\end{equation}

\noindent where $\tau$ is the softmax temperature. $g_0$ and $g_1$ are i.i.d. samples drawn from the standard Gumbel distribution G(0,1). Taking $g_0$ as an example, sampling a value $\gamma$ from the standard uniform distribution U(0,1) and then taking $-\log(-\log \gamma)$ as the final sample value $g_0$.

Based on the above reparameterization process with Gumbel-Softmax, we update the embedding $\mathbf{v}_i$ of the item $v_i$ with corresponding selection factor $\hat{d}_i$ to decrease noise interference in original behavior sequence, as shown in Eq. \ref{update item}.

\vspace{-7pt}
\begin{equation}
\begin{aligned}
\hat{\mathbf{v}}_i = \hat{d}_i\mathbf{v}_i
\end{aligned}
\label{update item}
\end{equation}

\subsection{Situational Feature Encoder}
% To capture the semantics of user behavior and explore user interest under various behavior situations, we devise the SRE (situation representation enhancing) module, which generates personalized parameters for distinct behavior situations at the situational feature level, and then enhances item representations by incorporating rich behavior situation semantics.
%To refine the embeddings of situational features, we devise the Situational Feature Encoder (SFE), which parameterizes the embeddings of situational features, with consideration of the commonality and difference between user interest under different values of the same situational feature, and then approximates interaction between item ID and some situational feature.

%For one behavior of a user interacting with an item, Situational Feature Encoder (SFE) generates the embedding for each situational feature of the behavior by combining commonalities and differences of the situational feature, and parameterizes the embeddings of situational features, approximating interaction between item ID and some situational feature.

For one behavior of a user interacting with an item, Situational Feature Encoder (SFE) generates the embeddings for situational features of the behavior by combining commonalities and differences in the situational features of the same type, and parameterizes the embeddings of situational features in the form of MLP, approximating the interaction between item ID and each situational feature.

Concretely, for the situational feature $f^k,1\le k\le n$, we generate the general vector $\mathbf{f}^k$ for this feature (e.g., period of the week) and the specific vector list $\{\mathbf{f}^{k, t}\}_{t=1}^{m_k}$ for all values of this situational feature (e.g., weekday or weekend), where $m_{k}$ is the number of values of the situational feature $f^k$ (e.g., $m_k=2$ for the above example). We think that the general vector and the specific vector list describe the commonalities and differences in the situational features of the same type, respectively.
%between user interest under different values of the same situational feature.
% thus improving the personalized situation awareness of the model, enhancing feature interaction, and avoiding poor generalization. 

Specifically, for situational feature $f_i^k$, we adopt its embedding $\mathbf{f}_i^k$ as the specific vector. Then, taking into consideration the candidate information, SFE fuses the general vector $\mathbf{f}^k$ and the specific vector $\mathbf{f}_i^{k}$ as shown in Eq. \ref{mlpfuse_1} and Eq. \ref{mlpfuse_2}.

\vspace{-7pt}
\begin{equation}
\begin{aligned}
\operatorname{gate}_{i,k} = \operatorname{sigmoid}\left(\operatorname{MLP}((\mathbf{v}_c \| \mathbf{f}_c^{k}) \otimes (\hat{\mathbf{v}}_i \| \mathbf{f}_i^{k}))\right)
\end{aligned}
\label{mlpfuse_1}
\end{equation}

\vspace{-7pt}
\begin{equation}
\begin{aligned}
\hat{\mathbf{{f}}}_{i}^k = \operatorname{gate}_{i,k} \mathbf{f}^{k}_{i} + \left(1-\operatorname{gate}_{i,k} \right) \mathbf{f}^k
\end{aligned}
\label{mlpfuse_2}
\end{equation}

\noindent where $\|$ indicates the concatenation operation and it is easy to find that the dimension of  $\hat{\mathbf{{f}}}_{i}^k$ is equal to $\mathbf{f}_i^k$, that is, $d_1$.

Then, $\hat{\mathbf{{f}}}_{i}^k$ is reshaped and split into the weight matrix and bias vector as the parameters of a micro-MLP named $\text{MLP}_i^k$ w.r.t. $f_i^k$. This process can be formularized as follows.

\vspace{-7pt}
\begin{equation}
\begin{aligned}
\left(w_{i,j}^k \| b_{i,j}^k\right)\|_{j=0}^{D-1}=\hat{\mathbf{{f}}}_{i}^k
\end{aligned}
\label{reshape-1}
\end{equation}

\vspace{-7pt}
\begin{equation}
\begin{aligned}
\sum_{j=0}^{D-1}\left(\left|w_{i,j}^k\right|+\left|b_{i,j}^k\right|\right)=\left|\hat{\mathbf{{f}}}_{i}^k\right|=d_1
\end{aligned}
\label{reshape-2}
\end{equation}

\noindent where $w_{i,j}^k$ and $b_{i,j}^k$ denote the weight and bias of $j$-th layer of $\text{MLP}_i^k$, $D$ determines the depth of the micro-MLP and $\left|\cdot\right|$ gets the size of the variables. Then, we feed $\hat{\mathbf{v}}_i$ into $\text{MLP}_i^k$ as Eq. \ref{feed-1} to refine the embedding of the situational feature $f_i^k$.

\vspace{-7pt}
\begin{equation}
\begin{aligned}
\mathbf{v}_{i,k} = \text{MLP}_i^k(\hat{\mathbf{v}}_i)
\end{aligned}
\label{feed-1}
\end{equation}

Following the above process, we can obtain the refined situational features representations $\{\mathbf{v}_{i,k}\}_{k=1}^{n}$ for each historical behavior $s_i$ and $\{\mathbf{v}_{c,k}\}_{k=1}^{n}$ for the target behavior (i.e., the click on the candidate $v_c$).

\subsection{Correlation Fusion Module}
In this section, we devise the Correlation Fusion Module (CFM) to coherently learn tri-directional correlations. Specifically, the first direction is along the behavior sequence. The information in the second direction refers to the relationships among situational features of the same behavior. Moreover, the third direction information describes the correlations among different dimensions (i.e., channels) of the situational feature embedding, where distinct channels contains various latent semantics.

Concretely, CFM consists of stacked $M$ layers of the identical block. Each block is equipped with three MLP-based mixers, i.e., a $\textit{behavior-mixer}$, a $\textit{channel-mixer}$, and a $\textit{feature-mixer}$, capturing the cross-behavior, cross-channel and cross-feature (i.e., cross-situational feature) correlations, respectively. Within each block, we first apply the $\textit{behavior-mixer}$ on the behavior sequence and then utilize $\textit{channel-mixer}$ to learn the latent semantics of the embedding for each situational feature. Then, $\textit{feature-mixer}$ is adopted to learn intrinsic correlations among all situational features of the same behavior. Specifically, the proposed CFM can learn tri-directional information without introducing the position information, unlike the series of methods based on Transformer, since CFM is an MLP-based method and the MLP is inherently sensitive to the position. It is worth noting that the parameters of each block are shared between $M$ layers to lighten the burden of model training and online inference.

\subsubsection{Preprocessing}
\ 
\newline
Based on the refined representation $\mathbf{v}_{i,k} \in \mathbb{R}^{d_2}$ of the situational feature $f_i^k$ obtained by the SFE module, we stack $\{\mathbf{v}_{i,k}\}_{i=1}^L$ and generate a two-dimensional matrix $\mathbf{V}^k \in \mathbb{R}^{L\times d_2}$ corresponding to $f^k$. Following this line, we further stack $\{\mathbf{V}^k\}_{k=1}^{n}$ to construct a three-dimensional matrix $\mathbf{V}\in \mathbb{R}^{n\times L \times d_2}$. 
% The $d_2$-axis of $\mathbf{V}$ contains channel information of a situational feature embedding, each column implies the sequential dependency, and the correlations among features are included in cross-feature layers along the $n$-axis. 

\subsubsection{Behavior-Mixer}
\ 
\newline
The $\textit{behavior-mixer}$ aims to capture the correlations among behaviors in the user historical behavior sequence. As aforementioned, we first apply the $\textit{behavior-mixer}$ and the $\textit{channel-mixer}$ successively for each situational feature and then adopt the $\textit{feature-mixer}$. Therefore, we detail the $\textit{behavior-mixer}$ based on the two-dimensional matrix $\mathbf{V}^k \in \mathbb{R}^{L \times d_2}$ corresponding to the situational feature $f^k$. 

The simplest implementation of $\textit{behavior-mixer}$ is feeding the column vector of $\mathbf{V}^k$ into a $\operatorname{MLP}$ block, which contains two fully-connected layers with a nonlinear activation function, and then outputting an embedding with the same dimension as the input. For easy understanding, the above process is essentially equivalent to transposing the matrix $\mathbf{V}^k$ and then performing the transformation on the row vector of the ${\mathbf{V}^k}^{\top}$ as shown in Fig. \ref{model architecture}. The $\textit{behavior-mixer}$ maps from $\mathbb{R}^L$ to $\mathbb{R}^L$ and is shared across all columns of $\mathbf{V}^k$. Formally, the output $\mathbf{V}^{k,\operatorname{Beh}}$ of the $\textit{behavior-mixer}$ for the situational feature $f^k$ can be represented as Eq. \ref{seq-mixer}.

\vspace{-7pt}
\begin{equation}
\begin{aligned}
\mathbf{V}_{*,\alpha}^{k,\operatorname{Beh}} = \mathbf{V}_{*,\alpha}^k + \mathbf{W}_2~\sigma\left(\mathbf{W}_1~{\operatorname{LayerNorm}\left(\mathbf{V}^k \right)}_{*,\alpha}\right), \quad \quad \quad \operatorname{for} ~ \alpha=1,\dots,d_2
\end{aligned}
\label{seq-mixer}
\end{equation}

\noindent where $\mathbf{V}_{*,\alpha}^{k,\operatorname{Beh}}$ is the $\alpha$-th column of $\mathbf{V}^{k,\operatorname{Beh}}$ and $\sigma$ is an activation function named $\operatorname{GELU}$ \cite{hendrycks2016gelu}. $\mathbf{W}_1 \in \mathbb{R}^{D_L \times L}$ and $\mathbf{W}_2 \in \mathbb{R}^{L \times D_L}$ denote two trainable matrices corresponding to two fully-connected layers, respectively. $D_L$ is the tunable hidden size of the $\textit{behavior-mixer}$. Aside from the MLP layers, we further employ other standard architectural components, i.e., layer normalization \cite{ba2016layernorm} and residual connection \cite{he2016residual}. 
It can be easily found that the $\textit{behavior-mixer}$ is sensitive to the order of behaviors in a sequence by adopting the MLP-based architecture, which frees the $\textit{behavior-mixer}$ from the position embedding.
% It can be easily found that the $\textit{behavior-mixer}$ is sensitive to the sequential order of the input,  which helps the CFM get rid of the position embedding. 

Although the above process can capture the correlations among all behaviors, there exist three critical issues. First, this method indiscriminately feeds the entire behavior sequence of length $L$ into the same $\operatorname{MLP}$ block, which may lead to noise interference. Second, the intrinsic and complex relationships of the behavior subsequence, consisting of adjacent behaviors within a certain range, are neglected by the above method. Third, the behavior subsequence is not strictly independent of each other and the latent correlations among them should be modeled by some mechanism.

To deal with the above three issues, we optimize the $\textit{behavior-mixer}~$ with a partition-and-fusion pattern. To materialize this idea, we divide the entire sequence of length $L$ into three types of subsequences according to different partition strategies. Then, we derive three distinct $\textit{behavior-mixer}$ to capture the correlations implied in the above three types of subsequences, respectively. We elaborate on these three $\textit{behavior-mixers}$ as follows.

\noindent $\textbf{Adjacent Behavior-Mixer}$ sets the window size as $L_w$, where $L$ is divisible by $L_w$, and then divides the entire sequence into $\frac{L}{L_w}$ segments from the beginning, i.e., each segment consists of $L_w$ behaviors. For the $\varphi$-th segment, where $\varphi \in \{1,\dots,\frac{L}{L_w}\}$, the consecutive $L_w$ behaviors are mapped by the $\operatorname{MLP}$ block, which has a nonlinear activation function and two fully connected layers denoted as $\mathbf{W}_1^{\varphi} \in \mathbb{R}^{D_{L_w}\times L_w}$ and $\mathbf{W}_2^{\varphi} \in \mathbb{R}^{L_w \times D_{L_w}}$, respectively. It is worth noting that the $\operatorname{MLP}$ blocks adopted by different segments are shared. The calculation result of the $\textit{adjacent behavior-mixer}$ for the $\varphi$-th segment is denoted as $\mathbf{V}^{k,\text{Adj}}\langle \varphi \rangle$, which can be formalized as follows.

\vspace{-7pt}
\begin{equation}
\begin{aligned}
\mathbf{V}^{k,\text{Adj}}_{*,\alpha}\langle \varphi \rangle = \mathbf{V}^{k,\text{Div}}_{*,\alpha}\langle \varphi \rangle + \mathbf{W}_2^{\varphi}~\sigma\left(\mathbf{W}_1^{\varphi}~{\operatorname{LayerNorm}\left(\mathbf{V}^{k,\text{Div}}\langle \varphi \rangle \right)}_{*,\alpha}\right), \quad \quad \quad \operatorname{for} ~ \alpha=1,\dots,d_2
\end{aligned}
\label{adj-seq-mixer}
\end{equation}

\noindent where $\mathbf{V}^{k,\text{Adj}}\langle \varphi \rangle \in \mathbb{R}^{L_w \times d_2}$ and $\mathbf{V}^{k,\text{Div}}\langle \varphi \rangle$ refers to the $\varphi$-th segment obtained by dividing the matrix $\mathbf{V}^k \in \mathbb{R}^{L\times d_2}$ along the $L$-axis. Then, we concat all of $\mathbf{V}^{k,\text{Adj}}_{*,\alpha}\langle \varphi \rangle \in \mathbb{R}^{L_w \times d_2}$ with $\varphi$ from $1$ to $\frac{L}{L_w}$ to derive the final output of the $\textit{adjacent behavior-mixer}$ as shown in Eq. \ref{cat-adj}. 

\vspace{-7pt}
\begin{equation}
\begin{aligned}
\mathbf{V}^{k,\text{Adj}}_{*,\alpha} = \left(\mathbf{V}^{k,\text{Adj}}_{*,\alpha}\langle \varphi \rangle\right)\|_{\varphi=1}^{\frac{L}{L_w}}, \quad \quad \quad \operatorname{for} ~ \alpha=1,\dots,d_2
\end{aligned}
\label{cat-adj}
\end{equation}

\noindent  $\textbf{Dilated Behavior-Mixer}$ is devised for capturing long-range correlations. Specifically, inspired by the dilated convolution, $\textit{dilated behavior-mixer}$ gathers behaviors from the entire sequence in intervals of $\frac{L}{L_w}$ and generate $\frac{L}{L_w}$ segments. Apparently,  each segment is composed of $L_w$ inconsecutive behaviors. The $\textit{dilated behavior-mixer}$ concentrates on learning the long-term dependency while disregarding the local patterns across adjacent behaviors, which is complementary to the $\textit{adjacent behavior-mixer}$. For $L_w$ behaviors in the $\varphi^{\prime}$-th segment, where $\varphi^{\prime} \in \{1,\dots, \frac{L}{L_w}\}$, we feed them into a $\operatorname{MLP}$ block with the same structure as that of the $\textit{adjacent behavior-mixer}$. Similarly, the $\operatorname{MLP}$ blocks are shared among different segments. We implement the $\textit{dilated behavior-mixer}$ as Eq. \ref{dil-seq-mixer}.

\vspace{-7pt}
\begin{equation}
\begin{aligned}
\mathbf{V}^{k,\text{Dil}}_{*,\alpha}\langle \varphi^{\prime} \rangle = \mathbf{V}^{k,\text{Gat}}_{*,\alpha}\langle \varphi^{\prime} \rangle + \mathbf{W}_2^{\varphi^{\prime}}~\sigma\left(\mathbf{W}_1^{\varphi^{\prime}}~{\operatorname{LayerNorm}\left(\mathbf{V}^{k,\text{Gat}}\langle \varphi^{\prime} \rangle \right)}_{*,\alpha}\right), \quad \quad \quad \operatorname{for} ~ \alpha=1,\dots,d_2
\end{aligned}
\label{dil-seq-mixer}
\end{equation}

\noindent where $\mathbf{V}^{k,\text{Dil}}\langle \varphi^{\prime} \rangle \in \mathbb{R}^{L_w \times d_2}$ is the output of $\textit{dilated behavior-mixer}$ with respect to the $\varphi^{\prime}$-th segment, and $\mathbf{V}^{k,\text{Gat}}\langle \varphi^{\prime} \rangle$ refers to the $\varphi^{\prime}$-th segment generated by gathering behaviors from $\mathbf{V}^k \in \mathbb{R}^{L\times d_2}$ along the $L$-axis in intervals of $\frac{L}{L_w}$ , starting with the $\varphi^{\prime}$-th row of $\mathbf{V}^k$. Then, we concat all of $\mathbf{V}^{k,\text{Dil}}_{*,\alpha}\langle \varphi^{\prime} \rangle \in \mathbb{R}^{L_w \times d_2}$ with $\varphi^{\prime}$ from $1$ to $\frac{L}{L_w}$ as Eq. \ref{concat-dil}.

\vspace{-7pt}
\begin{equation}
\begin{aligned}
\mathbf{V}^{k,\text{Dil}}_{*,\alpha} = \left(\mathbf{V}^{k,\text{Dil}}_{*,\alpha}\langle \varphi^{\prime} \rangle\right)\|_{\varphi^{\prime}=1}^{\frac{L}{L_w}}, \quad \quad \quad \operatorname{for} ~ \alpha=1,\dots,d_2
\end{aligned}
\label{concat-dil}
\end{equation}

\noindent $\textbf{Shifted Behavior-Mixer}$ aims to learn the correlations of two adjacent segments, which is another supplementary operation of the $\textit{adjacent behavior-mixer}$. Specifically, we shift the partition of segments in the $\textit{adjacent behavior-mixer}$ by an offset of $\lfloor \frac{L_w}{2} \rfloor$. For several behaviors near the  beginning and end of the sequence, since they are not covered by the shifted window, we concatenate them into $\mathbf{V}^{k,\text{Remain}} \in \mathbb{R}^{L_w \times d_2}$ and skip them when performing the $\operatorname{MLP}$ mapping. Apparently, shift-window operation generates $\frac{L}{L_w}-1$ segments. Then, for the $\varphi^{\prime\prime}$-th segments, where $\varphi^{\prime\prime} \in \{1,\dots,\frac{L}{L_w}-1\}$, we feed the consecutive $L_w$ behaviors in the $\varphi^{\prime\prime}$-th segments into a specific $\operatorname{MLP}$ block, which is shared between different segments. The $\textit{shifted behavior-mixer}$ can be formalized as Eq. \ref{shift-seq-mixer}.

\vspace{-7pt}
\begin{equation}
\begin{aligned}
\mathbf{V}^{k,\text{Shi}^{\prime}}_{*,\alpha}\langle \varphi^{\prime\prime} \rangle = \mathbf{V}^{k,\text{Sft}}_{*,\alpha}\langle \varphi^{\prime\prime} \rangle + \mathbf{W}_2^{\varphi^{\prime\prime}}~\sigma\left(\mathbf{W}_1^{\varphi^{\prime\prime}}~{\operatorname{LayerNorm}\left(\mathbf{V}^{k,\text{Sft}}\langle \varphi^{\prime\prime} \rangle \right)}_{*,\alpha}\right), \quad \quad \quad \operatorname{for} ~ \alpha=1,\dots,d_2
\end{aligned}
\label{shift-seq-mixer}
\end{equation}

\noindent where $\mathbf{V}^{k,\text{Shi}^{\prime}}\langle \varphi^{\prime\prime} \rangle \in \mathbb{R}^{L_w \times d_2}$ is the output of $\textit{shifted behavior-mixer}$ with respect to the $\varphi^{\prime\prime}$-th segment, and $\mathbf{V}^{k,\text{Sft}}\langle \varphi^{\prime\prime} \rangle$ refers to the $\varphi^{\prime\prime}$-th segment obtained by shifting the $\mathbf{V}^{k,\text{Div}}\langle \varphi^{\prime\prime} \rangle$ by an offset of $\lfloor \frac{L_w}{2} \rfloor$  along the $L$-axis. Then, we concat all of $\mathbf{V}^{k,\text{Shi}^{\prime}}_{*,\alpha}\langle \varphi^{\prime\prime} \rangle \in \mathbb{R}^{L_w \times d_2}$ with $\varphi^{\prime\prime}$ from $1$ to $\frac{L}{L_w}-1$ and the remaining behaviors representations $\mathbf{V}^{k,\text{Remain}} \in \mathbb{R}^{L_w \times d_2}$ to generate the final output of the $\textit{shifted behavior-mixer}$ as shown in Eq. \ref{concat-sw}. 

\vspace{-7pt}
\begin{equation}
\begin{aligned}
\mathbf{V}^{k,\text{Shi}}_{*,\alpha} = \mathbf{V}^{k,\text{Remain}}_{*, \alpha} \| \left(\left(\mathbf{V}^{k,\text{Shi}^{\prime}}_{*,\alpha}\langle \varphi^{\prime\prime} \rangle\right)\|_{\varphi^{\prime\prime}=1}^{\frac{L}{L_w}-1}\right), \quad \quad \quad \operatorname{for} ~ \alpha=1,\dots,d_2
\end{aligned}
\label{concat-sw}
\end{equation}

Now, three types of complementary correlations among all behaviors have been captured by the $\textit{adjacent behavior-mixer}$, $\textit{dilated behavior-mixer}$ and $\textit{shifted behavior-mixer}$, respectively. To obtain $\mathbf{V}^{k,\operatorname{Beh}}$, which incorporates above three types of behavior correlations, we aggregate $\mathbf{V}^{k,\text{Adj}} \in \mathbb{R}^{L \times d_2}, \mathbf{V}^{k,\text{Dil}}  \in \mathbb{R}^{L \times d_2}$ and $\mathbf{V}^{k,\text{Shi}}\in \mathbb{R}^{L \times d_2}$ in a weight-adaptive manner as shown in Eq. \ref{fuse-seq-mixer}.

\vspace{-7pt}
\begin{equation}
\begin{aligned}
\mathbf{V}^{k,\operatorname{Beh}} = w_1 \mathbf{V}^{k,\text{Adj}} + w_2 \mathbf{V}^{k,\text{Dil}}+ w_3\mathbf{V}^{k,\text{Shi}}
\end{aligned}
\label{fuse-seq-mixer}
\end{equation}

\noindent where $w_1, w_2$ and $w_3$ are learnable weight parameters.

\subsubsection{Channel-Mixer}
\ 
\newline
Aside from the correlations among different behaviors, the internal relationships among different channels of the situational feature embedding deserve more attention. Specifically, the embedding of the situational feature is multi-dimensional and usually implies latent semantics on each dimension. Modeling the intrinsic relationships among these dimensions by some mechanism can further improve performance. 

After performing the behavior mixing, the correlations along the user behavior sequence have been injected into  $\mathbf{V}^{k,\operatorname{Beh}}$ for each situational feature $f^k$ as shown in Eq. \ref{fuse-seq-mixer}. Then, the $\textit{channel-mixer}$ takes the row vector of $\mathbf{V}^{k,\operatorname{Beh}} \in \mathbb{R}^{L \times d_2}$ as input, feeds it into a $\operatorname{MLP}$ block and maps from $\mathbb{R}^{d_2}$ to $\mathbb{R}^{d_2}$ as shown in Eq. \ref{dim-mixer}. All rows of $\mathbf{V}^{k,\operatorname{Beh}}$ share the $\textit{channel-mixer}$.

\vspace{-7pt}
\begin{equation}
\begin{aligned}
\mathbf{V}_{\beta,*}^{k,\operatorname{Cha}} = \mathbf{V}_{\beta,*}^{k,\operatorname{Beh}} + \left(\mathbf{W}_4~\sigma\left(\mathbf{W}_3~\left({\operatorname{LayerNorm}\left(\mathbf{V}^{k,\operatorname{Beh}} \right)}_{\beta,*}\right)^\mathrm{T}\right)\right)^\mathrm{T}, \quad \quad \quad \operatorname{for} ~ \beta=1,\dots, L
\end{aligned}
\label{dim-mixer}
\end{equation}

\noindent where $\mathbf{V}_{\beta,*}^{k,\operatorname{Cha}}$ is the $\beta$-th row of $\mathbf{V}^{k,\operatorname{Cha}}$, $\sigma$ is $\operatorname{GELU}$, $\mathbf{W}_3 \in \mathbb{R}^{D_{d_2} \times d_2}$ and $\mathbf{W}_4 \in \mathbb{R}^{d_2 \times D_{d_2}}$  denote two trainable matrices. $D_{d_2}$ is the tunable hidden size of the $\textit{channel-mixer}$. After Eq. \ref{dim-mixer}, the internal relationships among different channels have been incorporated into $\mathbf{V}^{k,\operatorname{Cha}}$.

\subsubsection{Feature-Mixer}
\ 
\newline
After behavior mixing and channel mixing, the cross-behavior and cross-channel correlations have been incorporated into the embedding of each situational feature. However, the inherent and complex relationships among all the situational features, which jointly portray the situation of the behavior, have not been effectively captured. Thus, we derive the $\textit{feature-mixer}$ to fuse the cross-feature correlation into the embedding of the situational feature. Concretely, for each situational feature $f^k$, the cross-behavior and cross-channel relationships have been injected into $\mathbf{V}^{k,\operatorname{Cha}}$ along the path $\mathbf{V}^k \rightarrow \mathbf{V}^{k,\operatorname{Beh}} \rightarrow \mathbf{V}^{k,\operatorname{Cha}}$. We re-express $\mathbf{V}^{k,\operatorname{Cha}}$ as $\widetilde{\mathbf{V}}^k$ and concat all $\widetilde{\mathbf{V}}^k \in \mathbb{R}^{L\times d_2}$, where $k=1,\dots,n$, corresponding to $n$ situational features, to generate a three-dimensional matrix $\widetilde{\mathbf{V}}\in \mathbb{R}^{n\times L \times d_2}$. We transpose it into $\widetilde{\mathbf{V}}\in \mathbb{R}^{L \times d_2 \times n}$ and feed $\widetilde{\mathbf{V}}^{i}\in \mathbb{R}^{d_2 \times n}$ for each behavior $s_i$ into the $\textit{feature-mixer}$ separately. The $\textit{feature-mixer}$ maps from $\mathbb{R}^{n}$ to $\mathbb{R}^{n}$ for each row vector of $\widetilde{\mathbf{V}}^{i}$ with a  shared $\operatorname{MLP}$ block, which has a similar marco architecture to the $\textit{channel-mixer}$. This process is shown in Eq. \ref{fea-mixer}.

\vspace{-7pt}
\begin{equation}
\begin{aligned}
\widetilde{\mathbf{V}}^{i, \operatorname{Fea}}_{\alpha,*} = \widetilde{\mathbf{V}}^{i}_{\alpha,*} + \left(\mathbf{W}_6~\sigma\left(\mathbf{W}_5~\left({\operatorname{LayerNorm}(\widetilde{\mathbf{V}}^{i} )}_{\alpha,*}\right)^\mathrm{T}\right)\right)^\mathrm{T}, \quad \quad \quad \operatorname{for} ~ \alpha=1,\dots, d_2
\end{aligned}
\label{fea-mixer}
\end{equation}

\noindent where $\sigma$ is $\operatorname{GELU}$, $\mathbf{W}_5 \in \mathbb{R}^{D_n \times n}$ and $\mathbf{W}_6 \in \mathbb{R}^{n \times D_n}$  denote two trainable matrices. $D_n$ is the tunable hidden size of the $\textit{feature-mixer}$. Since the $\textit{feature-mixer}$ is subsequent to the $\textit{behavior-mixer}$ and $\textit{channel-mixer}$, empowers the $\textit{feature-mixer}$ not only to model the cross-feature correlations but also to build a bridge for information communication of cross-behavior and cross-channel correlations among all situational features. Therefore, the $\textit{feature-mixer}$ coherently connects the tri-directional information.

\subsection{Situation Aggregation Module}

We have incorporated the tri-directional information, i.e., cross-behavior, cross-channel, and cross-feature correlations, into $\widetilde{\mathbf{V}}^{i, \operatorname{Fea}} \in \mathbb{R}^{d_2 \times n}$ for each historical behavior $s_i$ as shown in Eq. \ref{fea-mixer}. Specifically, we slice $\widetilde{\mathbf{V}}^{i, \operatorname{Fea}}$ along the $n$-axis and obtain the $\{\tilde{\mathbf{v}}_{i,k}\}_{k=1}^n$ corresponding to $n$ situational features of the historical behavior $s_i$. It should be noted that we do not apply the CFM module on the situational features of the target behavior, thus the embeddings $\{\mathbf{v}_{c,k}\}_{k=1}^n$ generated by the SFE module are maintained. Subsequently, we adopt two different methods to aggregate $\{\tilde{\mathbf{v}}_{i,k}\}_{k=1}^n$ and $\{\mathbf{v}_{c,k}\}_{k=1}^n$, respectively, thus generating the embeddings for the historical behavior $s_i$ and the target behavior, respectively.

For the target behavior, with the consideration of importance difference among distinct situational features, we aggregate $\{\mathbf{v}_{c,k}\}_{k=1}^n$ into the embedding $\mathbf{b}_c$ of the target behavior in a weight-adaptive way, as shown in Eq. \ref{fuse-candidate}. 

\vspace{-7pt}
\begin{equation}
\begin{aligned}
\mathbf{b}_c = \operatorname{MLP}(\mathbf{v}_c + \sum_{k=1}^n  w_{c,k}\mathbf{v}_{c,k})
\end{aligned}
\label{fuse-candidate}
\end{equation}

\noindent where $w_{c,k}$ is a trainable weight parameter. 

Further, we dynamically reweight the embeddings of situational features for each historical behavior $s_i$ with respect to the situational features of the target behavior, and then generate the embedding $\mathbf{b}_i$ of the historical behavior $s_i$ as follows.

\vspace{-7pt}
\begin{equation}
\begin{aligned}
w_{i,k} = \operatorname{MLP}(\tilde{\mathbf{v}}_{i,k} + \mathbf{v}_{c,k} \otimes \tilde{\mathbf{v}}_{i,k})
\end{aligned}
\label{fuse-item-1}
\end{equation}

\vspace{-7pt}
\begin{equation}
\begin{aligned}
w_{i,k}^{\prime}=\frac{\exp(w_{i,k})}{\sum_{k=1}^n \exp({w_{i,k})}}
\end{aligned}
\label{fuse-item-2}
\end{equation}

\vspace{-7pt}
\begin{equation}
\begin{aligned}
\mathbf{b}_i = \operatorname{MLP}(\mathbf{v}_i + \sum_{k=1}^n w_{i,k}^{\prime}\tilde{\mathbf{v}}_{i,k})
\end{aligned}
\label{fuse-item-3}
\end{equation}

Then, to adjust the contributions of historical behavior representations at the channel level, we devise a channel-adaptive unit and give up the method in DIN \cite{zhou2018din}, since the method in DIN simply calculates the similarity score at the granularity of the item. For each historical behavior embedding $\mathbf{b}_i$, the channel-adaptive factor $\mathbf{g}_i$, which is a vector with the same dimension as $\mathbf{b}_i$ rather than a scalar, is computed as Eq. \ref{channel-adaptor}.

\vspace{-7pt}
\begin{equation}
\begin{aligned}
\mathbf{g}_i=\operatorname{sigmoid}\left(\mathbf{W}_{g}(\mathbf{b}_i \| (\mathbf{b}_i\otimes \mathbf{b}_c) \| \mathbf{b}_c)\right)
\end{aligned}
\label{channel-adaptor}
\end{equation}

\noindent where $\mathbf{W}_{g} \in \mathbb{R}^{d_2 \times 3d_2}$ is a learnable weight matrix. Finally, the embedding $\mathbf{e}_s$ of the behavior sequence is obtained by the following aggregation based on $\mathbf{g}_i$ as Eq. \ref{preference}.

\vspace{-7pt}
\begin{equation}
\begin{aligned}
\mathbf{e}_s = \frac{1}{L}\sum_{i=1}^L\mathbf{g}_i\otimes\mathbf{b}_i
\end{aligned}
\label{preference}
\end{equation}

\subsection{Model Optimization}

Finally, the user embedding $\mathbf{u}$, the embedding $\mathbf{e}_s$ of the behavior sequence, and the embedding $\mathbf{b}_c$ of the target behavior (i.e., the click on the candidate), along with other feature embeddings denoted as $\mathbf{e}_o$, are concatenated and fed into an $\operatorname{MLP}$ to perform the CTR prediction, as shown in Eq. \ref{ctr}.

\vspace{-7pt}
\begin{equation}
\begin{aligned}
\hat{y}=\operatorname{sigmoid}\left(\operatorname{MLP}(\mathbf{u}~\| ~\mathbf{e}_s~\|~\mathbf{b}_c~\|~\mathbf{e}_o)\right)
\end{aligned}
\label{ctr}
\end{equation}

DSAIN is optimized by minimizing the following negative log-likelihood function defined as Eq. \ref{loss}.

\vspace{-7pt}
\begin{equation}
\begin{aligned}
\mathcal{L}=-\frac{1}{N} \sum_{i=1}^{N}\left( y_{i} \log\hat{y}_i + \left(1-y_{i}\right) \log (1-\hat{y}_i)\right)
\end{aligned}
\label{loss}
\end{equation}

\noindent where $N$ denotes the batch size, $y_i \in \{0,1\}$ is the label and $\hat{y}_i$ is the predicted CTR value.

\subsection{Complexity Analysis}
\noindent \textbf{Space Complexity.}
The learnable parameters in DSAIN are from the item embeddings $(O(|\mathcal{V}|d_2))$, situational features embeddings $O(nd_1)$, and parameters in the BDM$(O(nd_1d_2))$, CFM$(O(nL_wD_{L_w}+nd_2D_{d_2}+nD_n))$, and SAM$(O(d_2^2))$. There are no extra parameters in SFE. Specifically, $d_1$ can be represented as $\delta d_2$, where $\delta$ is a constant with a certain range. Furthermore, $L_w$, $D_{L_w}$, $D_{d_2}$ and $D_n$ are hyperparameters, usually set to relatively small values. Thus, the total number of parameters is $O(|\mathcal{V}|d_2 + n d_2^2)$, which is moderate compared to other classic and effective methods (e.g., $O(|\mathcal{V}|d+d^2)$ for FeedRec\cite{wu2022feedrec}) since $n$ is quite limited.

% \vspace{+1pt}
\noindent \textbf{Time Complexity.} 
The computation amount of DSAIN is composed of four parts: the BDM module accounts for $O(Lnd_1d_2)$, the SFE module accounts for $O(Ln(d_1+d_2+d_2^2))$, the CFM module accounts for $O(MLnd_2)$, and the SAM module accounts for $O(Ld_2^2)$. As aforementioned, $d_1$ can be represented as $\delta d_2$. Therefore, the total computational complexity is $O(Lnd_2^2+MLnd_2)$. Since $n$ and $M$ are relatively small, our model does not increase the computational cost, compared to other CTR models (e.g., $O(nd^2+n^2d)$ for FeedRec\cite{wu2022feedrec}). A convenient property in our model is that the computation of all items in the behavior sequence is fully parallelizable, which is amenable to GPU acceleration.

\section{Experiments}
In this section, we conduct extensive offline experiments on three real-world datasets and an online A/B test to answer the following research questions.

\begin{itemize}[leftmargin=*]
\item  \textbf{RQ1}: Does DSAIN outperform the existing CTR models on performance?
\item \textbf{RQ2}: How do four components in the DSAIN contribute to the recommendation performance?
% \item \textbf{RQ3}: What is the impact of adopting different combinations of situational features on model performance?
\item \textbf{RQ3}: How do the key hyper-parameters in DSAIN affect its performance?
\item \textbf{RQ4}: What is the online performance of DSAIN in a live production environment, e.g., the Meituan food delivery platform?
\end{itemize}

\subsection{Experimental Setup}
\subsubsection{Datasets.} We conduct offline experiments on the following three datasets.

\begin{itemize}[leftmargin=*]
\item  Taobao. It is a public dataset\footnote{https://tianchi.aliyun.com/dataset/56} collected from the display advertising system of Taobao. This dataset contains more than 26 million interaction records of 1.14 million users within 8 days. Each sample comprises five features: user ID, timestamp, behavior type, item brand ID, and category ID. It should be noted that this dataset does not include item ID but adopts brand ID as the alternative. We construct temporal features based on timestamp information, which together with the behavior type constitute the situational features of this dataset.  
\item Eleme. It is a public dataset\footnote{https://tianchi.aliyun.com/dataset/131047} constructed from logs of ele.me service. This dataset contains interactions within 8 days. Aside from the behavior type, this dataset naturally contains rich temporal features (e.g., the hour of day, weekday or weekend, etc.), which jointly form the situational features.
\item Meituan. It is an industrial dataset collected from the Meituan food delivery platform, which contains 1.3 billion interaction records of 130 million users within 30 days. This industrial dataset not only includes behavior type and temporal features (e.g., the period of the day corresponding to breakfast/lunch/dinner, and weekday/weekend, etc.) but also incorporates behavior-related spatial information, such as the page (e.g., search result page and recommendation list page) in which the behavior occurs.
\end{itemize}

The statistics of the three datasets are shown in Table \ref{tab:dataset}.

\begin{table}
  \caption{Statistics of datasets.}
  \label{tab:dataset}
  \begin{tabular}{lcccc}
    \toprule
    Datasets& \#Users& \#Items& \#Categories& \#Behaviors\\
    \midrule
    Taobao & 1,141,729 & 99,815 & 6,769 & 26,557,961\\
    Eleme  & 14,427,689 & 7,446,116 & 10 & 177,114,244\\
    Meituan & 130,648,310 & 14,054,691 & 184 & 1,331,247,488\\
  \bottomrule
\end{tabular}
\end{table}

\subsubsection{Comparison Models.} 
%We conduct comparative experiments with the following eight baselines, which are divided into three groups.
The following eight baselines are chosen for comparative experiments and they are divided into three groups.

Group I includes three models that treat behavior-related side information as common features.
\begin{itemize}[leftmargin=*]
\item $\textbf{DIN}$\cite{zhou2018din} employs a local activation unit to dynamically reweight users’ historical behaviors w.r.t. the candidate.
\item $\textbf{DIEN}$\cite{zhou2019dien} introduces an interest-evolving layer to capture the evolution of interest on the basis of DIN.
\item $\textbf{CAN}$\cite{bian2022can} disentangles representation learning and feature interaction modeling via a co-action unit, which is also equipped with multi-order enhancement and multi-level independence to enhance the feature interaction.
\end{itemize}

Group II includes two models that conduct modeling for the behavior type.
\begin{itemize}[leftmargin=*]
\item $\textbf{DMT}$\cite{gu2020dmt} exploits Deep Multifaceted Transformers to model users’ diverse behavior sequences, utilizes MMoE to jointly optimize multi-objectives and uses the Bias Deep Neural Networks to reduce the selection bias in implicit feedback.
\item $\textbf{FeedRec}$\cite{wu2022feedrec} gives a user modeling framework to incorporate various explicit and implicit user feedback to infer user interest.
\end{itemize}

Group III includes three models that focus on modeling for behavior temporal and/or spatial information.
\begin{itemize}[leftmargin=*]
\item $\textbf{CARCA}$\cite{rashed2022carca} captures the dynamic nature of the user profiles in terms of contextual features and item attributes via dedicated multi-head self-attention blocks.
\item $\textbf{Trans2D}$\cite{singer2022trans2D} devises a novel extended attention mechanism named attention2D to learn item-item, attribute-attribute and item-attribute patterns from sequences with multiple item attributes.
\item $\textbf{DIF-SR}$\cite{xie2022dif} decouples the behavior-related side information and the item embedding to prevent information entanglement.
\end{itemize}

\subsubsection{Evaluation Metrics.} 
We adopt two metrics, i.e., $\textbf{AUC}$ (Area Under ROC Curve) and $\textbf{Logloss}$, to evaluate our proposed model in offline experiments. A larger AUC indicates better recommendation performance, but Logloss performs the opposite. It is noteworthy that a small improvement in AUC is likely to lead to a significant increase in online CTR \cite{zhou2018din}. In online experiments, we adopt $\textbf{CTR}$ (Click-Through Rate), $\textbf{CPM}$ (Cost-Per-Mille) and $\textbf{GMV}$ (Gross Merchandise Volume) as evaluation metrics. 

\subsubsection{Implementation Details. } 
We implement DSAIN by TensorFlow. For all comparison models and our DSAIN model, we adopt Adam as the optimizer with the learning rate fixed to 0.001 and initialize the model parameters with normal distribution by setting the mean and standard deviation to 0 and 0.01, respectively. For our proposed DSAIN model, $L$, $L_w$, $n$, $d_1$, and $d_2$ are set to 300, 30, 5, 144, and 8, respectively. The temperature parameter $\tau$ in BDM and the layer depth in CFM are set to 1 and 4, respectively. For the hidden dimensions of three mixers, i.e., $D_{L_w}$ for the $\textit{behavior-mixer}$, $D_{d_2}$ for the $\textit{channel-mixer}$, and $D_n$ for the $\textit{feature-mixer}$, in CFM are set to 10, 16, 8, respectively.

\begin{table*}
\setlength\tabcolsep{4pt}
  \caption{Performance comparison. The best result and the second-best result in each row are in bold and underlined, respectively.}
  \label{tab:performance}
  \begin{tabular}{llccccccccccc}
    \toprule[1pt]
     \multirow{2}{*}{Dataset} & \multirow{2}{*}{Metric} & \multicolumn{3}{c}{Group I} & \multicolumn{1}{c}{} & \multicolumn{2}{c}{Group II} & \multicolumn{1}{c}{} & \multicolumn{3}{c}{Group III} & \multirow{2}{*}{DSAIN} \\
    \cline{3-5} 
    \cline{7-8}
    \cline{10-12} 
    % \multicolumn{1}{c}{}&
     & & DIN & DIEN & CAN & & DMT & FeedRec& & CARCA & Trans2D & DIF-SR &  \\
    \midrule
    \multirow{2}{*}{Taobao}
    & AUC & 0.6223 & 0.6241 & 0.6276 & & 0.6268 & 0.6312 & & 0.6289 & 0.6303 & \underline{0.6330} & \textbf{0.6452} $\pm$ 0.5\textperthousand \\
    & Logloss & 0.2622 & 0.2618 & 0.2607 & & 0.2611 & 0.2592 & & 0.2603 & 0.2599 & \underline{0.2588} & \textbf{0.2571} $\pm$ 0.1\textperthousand \\
    \midrule
    \multirow{2}{*}{Eleme}
    & AUC & 0.6368 & 0.6420 & 0.6450 & & 0.6435 & 0.6477 & & 0.6455 & 0.6464 & \underline{0.6502} & \textbf{0.6634} $\pm$ 0.4\textperthousand \\
    & Logloss & 0.1245 & 0.1198 & 0.1157 & & 0.1173 & 0.1143 & & 0.1152 & 0.1149 & \underline{0.1135} & \textbf{0.1116} $\pm$ 0.1\textperthousand \\
    \midrule
    \multirow{2}{*}{Meituan}
    & AUC & 0.6577 & 0.6612 & 0.6645 & & 0.6635 & 0.6715 & & 0.6683 & 0.6709 & \underline{0.6740} & \textbf{0.6823} $\pm$ 0.4\textperthousand \\
    & Logloss & 0.1922 & 0.1884 & 0.1855 & & 0.1861 & 0.1819 & & 0.1842 & 0.1825 & \underline{0.1799} & \textbf{0.1763} $\pm$ 0.1\textperthousand \\
    \bottomrule[1pt]
  \end{tabular}
\end{table*}

\subsection{Overall Performance (RQ1)}
Table \ref{tab:performance} lists the performance results of baselines and our DSAIN model on Taobao, Eleme, and Meituan datasets. All experiments are repeated three times. The best results of compared models and the average results of DSAIN are reported. From Table \ref{tab:performance}, we find performance trends of all the baselines are consistent on three datasets and have some observations as follows.

(1) FeedRec in Group II outperforms all models in Group I, and DIF-SR in Group III outperforms all the other models in Groups I, II, and III. This demonstrates the significance of fully utilizing the behavior-related side information. 

(2) DMT in Group II performs better than DIN and DIEN in Group I, while slightly worse than CAN in Group I. Moreover, CARCA and Trans2D in Group III surpass models in Group I and DMT in Group II, but are slightly worse than FeedRec in Group II. This phenomenon inspires us that the improvement of model performance does not merely depend on utilizing richer information of interaction and the model architecture imposes a significant effect on the performance.

(3) Our DSAIN consistently outperforms all baseline models and achieves state-of-the-art performance on all datasets. We think, compared to DSAIN, all competitors either underutilize the behavior-related side information, or neglect to capture multi-directional correlations implied in the user behavior sequence, especially the dependencies among different situational features, thus suffering the degradation of model performance.

\subsection{Ablation Study (RQ2)}
In this section, we perform the ablation study on the Meituan dataset to investigate the different impacts of four modules on the performance of DSAIN. 

\subsubsection{Effect of Behavior Denoising Module}
In this section, we investigate the impact of the Behavior Denoising Module (BDM) in DSAIN. We consider the following variants: (1) $\operatorname{DSAIN}^*_1$: it simplifies the $\operatorname{DSAIN}$ by dropping BDM, and performs nothing on the original user historical behavior sequence and then directly feeds it into the model; (2) $\operatorname{DSAIN}^*_2$: it keeps a specified number of exposed items before each clicked item and discards the remaining exposed items, thereby forming a new sequence, which is the subsequence of the original user historical behavior sequence. Specifically, $\operatorname{DSAIN}^*_2(x)$ sets the number of reserved exposures to $x$, where $x \in \{2, 4, 8, 12, 16, 20\}$. 

As shown in Table \ref{tab:ablation-BDM}, (1) compared to $\operatorname{DSAIN}^*_1$ and $\operatorname{DSAIN}^*_2(\cdot)$, $\operatorname{DSAIN}$ consistently fosters the improvement of recommendation performance, which demonstrates the effectiveness of BDM. (2) We find that the model performance is the worst when only keeping two exposed items before each clicked item. This may be because the historical behavior sequence considered in this paper contains a large number of exposed items while clicked items are relatively sparse. Too much valuable information is discarded when almost all exposures are dropped. Therefore, $\operatorname{DSAIN}$ performs better when $x$ increases from 2 to 12. However, the model performance suffers a decrease as $x$ continues to increase from 12, which may be owing to the noise interference introduced by excessive exposures.

\begin{table}
  \caption{Performance of variants with different BDMs or without BDM.}
  \label{tab:ablation-BDM}
  \setlength{\tabcolsep}{1.2mm}
  \begin{tabular}{lcccccccc}
    \toprule
     & $\operatorname{DSAIN}^*_1$ & $\operatorname{DSAIN}^*_2(2)$ & $\operatorname{DSAIN}^*_2(4)$ & $\operatorname{DSAIN}^*_2(8)$ & $\operatorname{DSAIN}^*_2(12)$ & $\operatorname{DSAIN}^*_2(16)$ & $\operatorname{DSAIN}^*_2(20)$ & $\operatorname{DSAIN}$\\
    \midrule
    AUC & 0.6808 & 0.6778 & 0.6797 & 0.6808 & 0.6811 & 0.6804 & 0.6803 &  \textbf{0.6823}\\
    Logloss & 0.1765 & 0.1770 & 0.1767 & 0.1764 & 0.1764 & 0.1767 & 0.1765 & \textbf{0.1763} \\
  \bottomrule
\end{tabular}
\end{table}

\subsubsection{Effect of Situational Feature Encoder.}
To demonstrate the significance of the Situational Feature Encoder (SFE), we consider the following variants: (1) $\operatorname{DSAIN}^{\dagger}_1$: it simplifies $\operatorname{DSAIN}$ by replacing SFE with an average pooling over all $\hat{\mathbf{f}}_i^k, k=1,\dots,n,$ plus the concatenation of the result and the item embedding $\hat{\mathbf{v}}_i$; (2) $\operatorname{DSAIN}^{\dagger}_2$: it performs the average pooling over all $\hat{\mathbf{f}}_i^k, k=1,\dots,n,$ and obtain a vector $\bar{\mathbf{f}}_i^k$ , then splits $\bar{\mathbf{f}}_i^k$ into several vectors adopted as parameters of the weight matrix and the bias vector of a $\operatorname{micro-MLP}$, then feeds the item embedding $\hat{\mathbf{v}}_i$ into the micro-MLP; (3) $\operatorname{DSAIN}^{\dagger}_3$: it concatenates the item embedding $\hat{\mathbf{v}}_i$ with each situational feature embedding $\hat{\mathbf{f}}_i^k$ to generate $\mathbf{v}_{i,k}$; (4) $\operatorname{DSAIN}^{\dagger}_4$: it only uses the specific vector $\mathbf{f}_i^k$ and drops the general vector $\mathbf{f}^k$ to generate the embedding  $\hat{\mathbf{f}}_i^k$ for the situational feature $f_i^k$. 

The results are shown in Table \ref{tab:ablation-SFE}. We can infer from Table \ref{tab:ablation-SFE} that (1) the original $\operatorname{DSAIN}$ outperforms all four variants, which shows the effectiveness of SFE. (2) Compared to the original $\operatorname{DSAIN}$, the sharp performance degradation of $\operatorname{DSAIN}^{\dagger}_1$ and $\operatorname{DSAIN}^{\dagger}_2$ indicates that fusing situational features prematurely may lead to information distraction and deteriorate the model performance. In this case, even with a more complex interaction scheme, such as an $\operatorname{micro-MLP}$ structure in $\operatorname{DSAIN}^{\dagger}_2$, the performance improvement is also very limited in comparison to $\operatorname{DSAIN}^{\dagger}_1$, which highlights the necessity of considering each situational feature separately. (3) Compared to $\operatorname{DSAIN}$, $\operatorname{DSAIN}^{\dagger}_3$ drops the $\operatorname{micro-MLP}$ and suffers poor performance, which illustrates that the $\operatorname{micro-MLP}$ can assist in achieving full-fledged feature interactions between item embedding and situational features. (4) Comparing $\operatorname{DSAIN}^{\dagger}_4$ to DSAIN, the decrease in AUC indicates that the commonalities of the situational features of the same type are successfully injected into the general vector in SFE.

% \vspace{-0.5em} 
\begin{table}[htbp]
\setlength\tabcolsep{3pt}%调列距
\centering
\begin{minipage}[t]{0.45\linewidth}
\centering
\caption{Performance of variants with different SFEs.}
\label{tab:ablation-SFE}
\begin{tabular}{lccccc}
    \toprule
    & $\operatorname{DSAIN}^{\dagger}_1$ & $\operatorname{DSAIN}^{\dagger}_2$ & $\operatorname{DSAIN}^{\dagger}_3$ & $\operatorname{DSAIN}^{\dagger}_4$ & $\operatorname{DSAIN}$\\
    \midrule
    AUC & 0.6790 & 0.6795 & 0.6801 & 0.6810 & \textbf{0.6823}\\
    Logloss & 0.1766 & 0.1766 & 0.1765 & 0.1764 & \textbf{0.1763}\\
  \bottomrule
\end{tabular}
\end{minipage}
\hfill
\begin{minipage}[t]{0.45\linewidth}
\centering
\caption{Performance of variants with different SAMs.}
\label{tab:ablation-SAM}
\begin{tabular}{lcccc}
    \toprule
    & $\operatorname{DSAIN}^{\diamond}_1$ & $\operatorname{DSAIN}^{\diamond}_2$ &
    $\operatorname{DSAIN}^{\diamond}_3$ &
    $\operatorname{DSAIN}$\\
    \midrule
    AUC & 0.6805 & 0.6800 & 0.6812 & \textbf{0.6823}\\
    Logloss & 0.1765 & 0.1767 & 0.1764 & \textbf{0.1763}\\
  \bottomrule
\end{tabular}
\end{minipage}
\end{table}
\vspace{-1.5em}

\subsubsection{Effect of Correlation Fusion Module}
To validate the importance of the Correlation Fusion Module (CFM), we design seven variants as follows: (1) $\operatorname{DSAIN}^{\ddagger}_1$: it only performs behavior mixing; (2) $\operatorname{DSAIN}^{\ddagger}_2$: it only performs channel mixing; (3) $\operatorname{DSAIN}^{\ddagger}_3$: it only performs feature mixing; (4) $\operatorname{DSAIN}^{\ddagger}_4$: it removes the $\textit{feature-mixer}$ in CFM, i.e., performs behavior mixing and channel mixing; (5) $\operatorname{DSAIN}^{\ddagger}_5$: it simplifies CFM by removing the $\textit{channel-mixer}$; (6) $\operatorname{DSAIN}^{\ddagger}_6$: it simplifies CFM by removing the $\textit{behavior-mixer}$; (7) $\operatorname{DSAIN}^{\ddagger}_7$: it removes the nonlinear activation function $\operatorname{GELU}$ from the $\operatorname{MLP}$ block used in three mixers. The results are shown in Table \ref{tab:ablation-CFM}. 

From Table \ref{tab:ablation-CFM}, we can observe that (1) all seven variants suffer a decline in AUC and an increase in Logloss, compared to the original DSAIN. (2) Comparison between the variants with only a single mixer (i.e., $\operatorname{DSAIN}^{\ddagger}_1$, $\operatorname{DSAIN}^{\ddagger}_2$, $\operatorname{DSAIN}^{\ddagger}_3$) and the original $\operatorname{DSAIN}$ shows that the $\textit{behavior-mixer}$ and $\textit{feature-mixer}$ play more important roles than $\textit{channel-mixer}$, and the $\textit{behavior-mixer}$ is the most significant. (3) Comparing the variants with two mixers (i.e., $\operatorname{DSAIN}^{\ddagger}_4$, $\operatorname{DSAIN}^{\ddagger}_5$, $\operatorname{DSAIN}^{\ddagger}_6$) and the original $\operatorname{DSAIN}$, we find that $\operatorname{DSAIN}^{\ddagger}_6$ suffers most drastic performance degradation, which shows the $\textit{behavior-mixer}$ has a significant effect on model performance. Moreover, the AUC decrease of $\operatorname{DSAIN}^{\ddagger}_4$ by more than two thousand points compared to the original $\operatorname{DSAIN}$ illustrates that the $\textit{feature-mixer}$ coherently connects the tri-directional information, which fosters the improvement of model performance. (4) Performance decrease of $\operatorname{DSAIN}^{\ddagger}_7$ compared to the original $\operatorname{DSAIN}$ demonstrates the necessity of introducing nonlinearity into the CFM module by adopting the activation function $\operatorname{GELU}$. (5) Compared to using $\textit{behavior-mixer}$ or $\textit{feature-mixer}$ alone, further introducing the $\textit{channel-mixer}$ consistently leads to performance degradation, which may be due to information entanglement. Nevertheless, comparison between $\operatorname{DSAIN}^{\ddagger}_5$ and the original $\operatorname{DSAIN}$ shows the effectiveness of $\textit{channel-mixer}$ in capturing latent semantics across multiple channels while using three mixers together. (6) $\operatorname{DSAIN}$ achieves the best performance by jointly leveraging three mixers, thus coherently capturing the tri-directional correlations, i.e., cross-behavior, cross-channel, and cross-feature correlations.

\begin{table}
  \caption{Performance of variants with different CFMs.}
  \label{tab:ablation-CFM}
  \begin{tabular}{ccccccccc}
    \toprule
     & $\operatorname{DSAIN}^{\ddagger}_1$ & $\operatorname{DSAIN}^{\ddagger}_2$ & $\operatorname{DSAIN}^{\ddagger}_3$ & $\operatorname{DSAIN}^{\ddagger}_4$ & $\operatorname{DSAIN}^{\ddagger}_5$ & $\operatorname{DSAIN}^{\ddagger}_6$ & $\operatorname{DSAIN}^{\ddagger}_7$ &  $\operatorname{DSAIN}$\\
    \midrule
    \textit{behavior-mixer} & \checkmark &  &  & \checkmark & \checkmark &  & \checkmark & \checkmark \\
    \textit{channel-mixer} &  & \checkmark &  & \checkmark &  & \checkmark & \checkmark & \checkmark \\
    \textit{feature-mixer} &  &  & \checkmark &  & \checkmark & \checkmark & \checkmark & \checkmark\\
    GELU & \checkmark & \checkmark & \checkmark & \checkmark & \checkmark & \checkmark &  & \checkmark \\
    \midrule
    AUC & 0.6805 & 0.6779 & 0.6793 & 0.6798 & 0.6812 & 0.6770 & 0.6806  & \textbf{0.6823}\\
    Logloss & 0.1765 & 0.1767 & 0.1767 & 0.1766 & 0.1764 & 0.1769 & 0.1765 & \textbf{0.1763}\\
  \bottomrule
\end{tabular}
\end{table}
\vspace{-10pt} %调整图片与下文的垂直距离

\subsubsection{Effect of Situation Aggregation Module.}
In this section, we investigate the impact of the Situation Aggregation Module (SAM) in DSAIN by considering the following variants: (1) $\operatorname{DSAIN}^{\diamond}_1$: it performs the average pooling over all $\tilde{\mathbf{v}}_{i,k}, k=1,\dots, n,$ to aggregate the embeddings of situational features of the same behavior; (2) $\operatorname{DSAIN}^{\diamond}_2$: it aggregates $\tilde{\mathbf{v}}_{i,k}, k=1,\dots, n$ in a weight-adaptive manner; (3) $\operatorname{DSAIN}^{\diamond}_3$: it calculates the correlation score between the target behavior (a click on the candidate) and each historical behavior to dynamically adjust contributions of historical behaviors. 

From Table \ref{tab:ablation-SAM}, we find that (1) both $\operatorname{DSAIN}^{\diamond}_1$ and $\operatorname{DSAIN}^{\diamond}_2$ suffer a decline in AUC compared to the original $\operatorname{DSAIN}$, which demonstrates the significance of making the utmost of the target behavior to guide the aggregation of embeddings of situational features of the same behavior; (2) The performance degradation of $\operatorname{DSAIN}^{\diamond}_3$ compared to $\operatorname{DSAIN}$ shows the effectiveness of the channel-adaptive unit in the SAM, which empowers DSAIN to modulate contributions at the dimensional level of behavioral embeddings, where the dimensional level is more fine-grained than item level.

\subsection{Hyperparameter Analysis (RQ3)}
\begin{figure}[htbp]
    \centering
    \begin{subfigure}[b]{0.3\textwidth}
        \includegraphics[width=\textwidth]{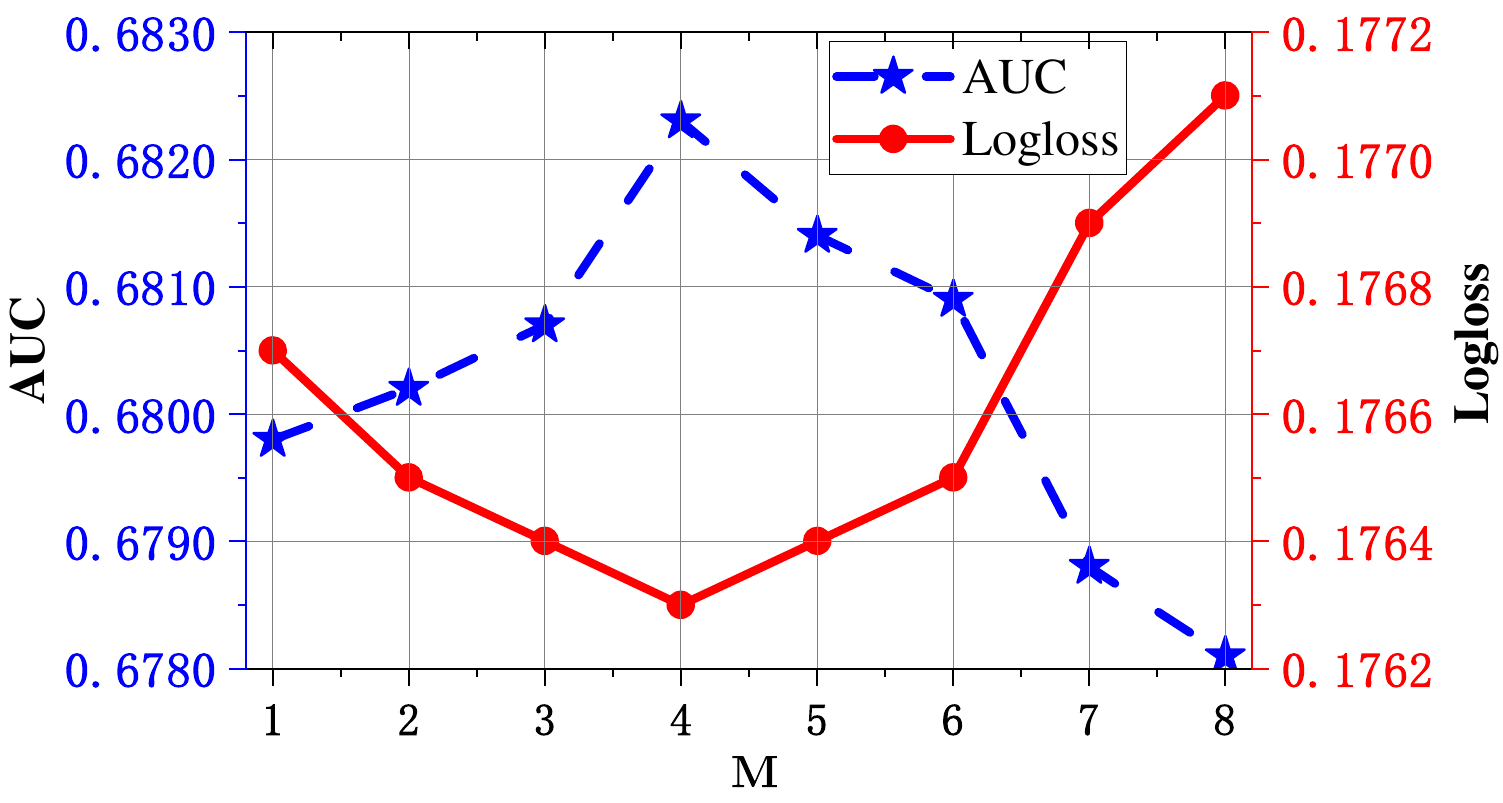}
        \caption{Layer depth $M$ in CFM.}
        \label{fig:sub1}
    \end{subfigure}
    \hspace{.1in}
    \begin{subfigure}[b]{0.3\textwidth}
        \includegraphics[width=\textwidth]{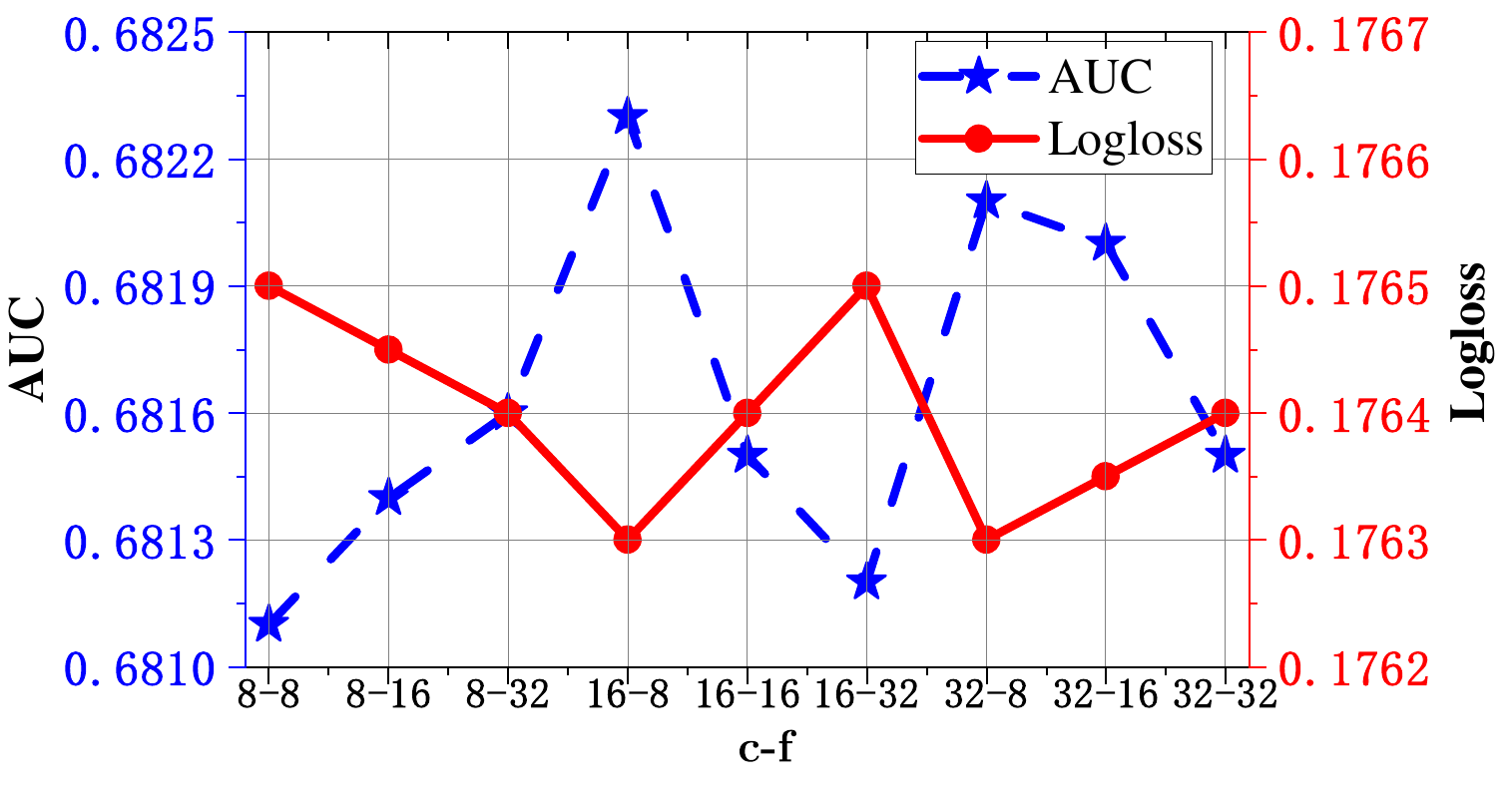}
        \caption{Hidden dimensions $\text{c-f}$ in CFM.}
        \label{fig:sub2}
    \end{subfigure}
    \hspace{.1in}
    \begin{subfigure}[b]{0.3\textwidth}
        \includegraphics[width=\textwidth]{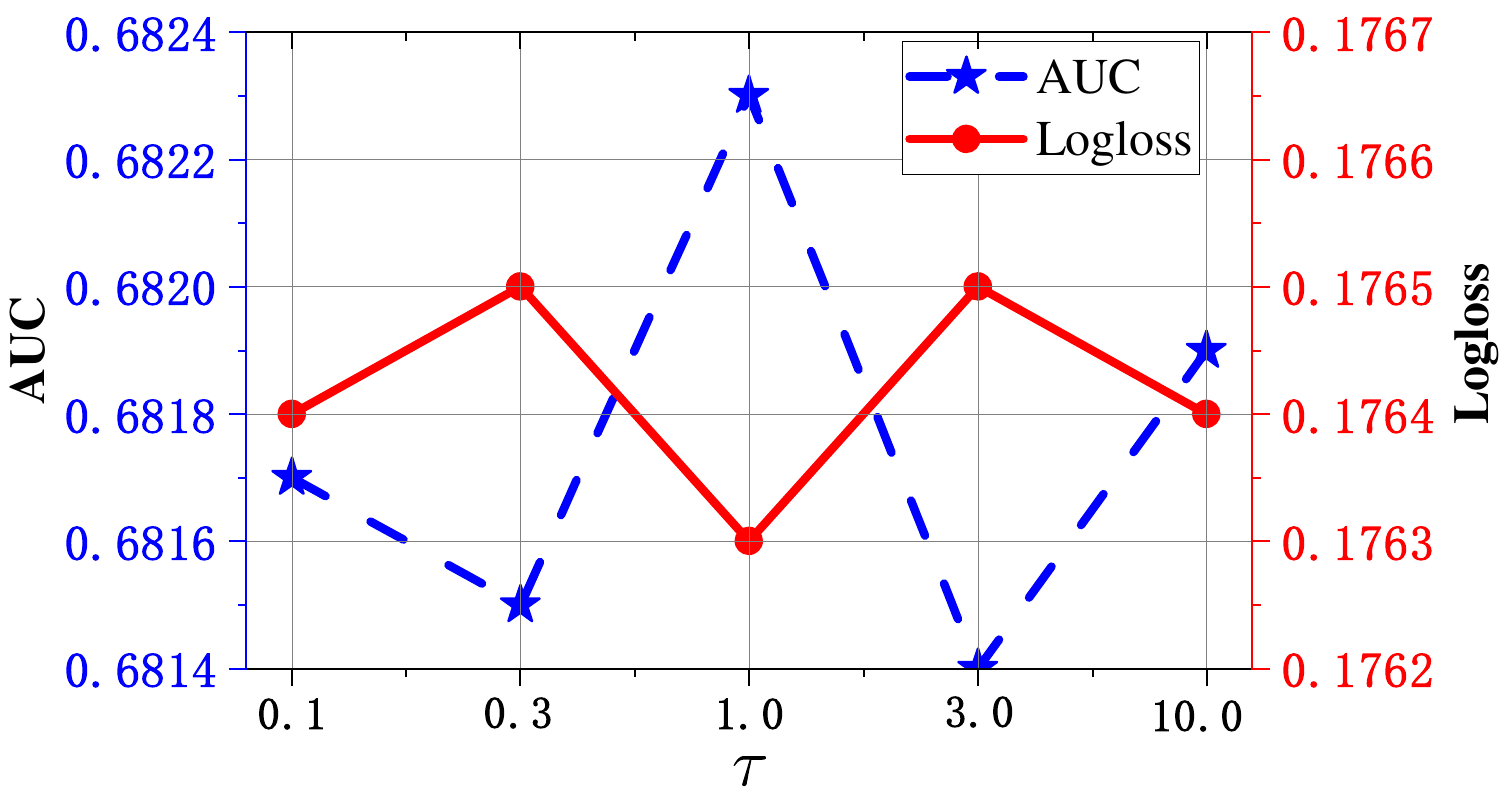}
        \caption{Temperature $\tau$ in BDM.}
        \label{fig:sub3}
    \end{subfigure}
    \caption{Influence of critical hyperparameters on the model performance.}
    \label{fig:parameters}
\end{figure}

In this section, we conduct the experiments to analyze the sensitivity of several critical hyperparameters in our model, including the layer depth $M$ in CFM (Correlation Fusion Module), the hidden dimensions of three mixers (i.e., $D_{L_w}$ of $\textit{behavior-mixer}$, $D_{d_2}$ of $\textit{channel-mixer}$ and $D_n$ of $\textit{feature-mixer}$) in CFM, and the temperature $\tau$ in BDM (Behavior Denoising Module). As mentioned in the methodology, CFM consists of stacked $M$ layers of the identical block. Thus, we first vary the layer depth $M$ from 1 to 8 to investigate its influence on model performance. As shown in Fig. \ref{fig:parameters} (a), the model performs better as $M$ increases from 1 to 4, but then worse as $M$ further increases. We consider the reason for this phenomenon is that stacking more layers moderately contributes to fully exploiting the tri-directional correlations, while excessive layers might also introduce noise interference and blur the representations of situational features. Therefore, we set $M$ to 4 according to the results.

% \vspace{-0.8cm} %调整图片与下文的垂直距离

Then, we evaluate the effect of the hidden dimensions of three mixers (i.e., $D_{L_w}$, $D_{d_2}$, and $D_n$) in CFM. Firstly, fixing $D_{d_2}$ to 16 and $D_n$ to 8, we tune the hidden dimension $D_{L_w}$ of the $\textit{behavior-mixer}$ in $\{2, 4, 8, 10, 12, 16, 20\}$ w.r.t. the sequence with a total length of 300 used in our DSAIN. The results present DSAIN performs best when $D_{L_w}=10$. Due to the limited space, we omit the visualization of the results. Then, we consider different combinations of hidden dimensions $D_{d_2}\in \{8, 16, 32\}$ of the $\textit{channel-mixer}$ and $D_n\in \{8, 16, 32\}$ of the $\textit{feature-mixer}$ with $D_{L_w}$ fixed to 10. Fig. \ref{fig:parameters} (b) shows the performance of DSAIN with different configurations of $D_{d_2}$ and $D_n$ denoted as "$\text{c-f}$" (e.g., "$\text{8-16}$" represents $D_{d_2}=8$ and $D_n=16$). From Fig. \ref{fig:parameters} (b), we find that $\operatorname{DSAIN}$ achieves the peak performance when $D_{d_2}=16$ and $D_n=8$. 
% Therefore, we set $D_{L_w}$, $D_{d_2}$, and $D_n$ to 10, 16, and 8, respectively.

Finally, we tune the temperature $\tau$ in $\{0.1, 0.3, 1.0, 3.0, 10.0\}$ to observe its influence on model performance. From Fig. \ref{fig:parameters} (c), we find that a moderate value of $\tau$ such as 1.0 is suitable for our DSAIN, while the performance is suboptimal when $\tau$ is either too small or too large.

\subsection{Online A/B Test (RQ4)}
We conduct the online A/B test by deploying our DSAIN to handle real 40\% traffic in the Meituan food delivery platform for seven days beginning on December 6, 2022, where the baseline model is the online-serving CTR model of the Meituan food delivery platform. Compared to the baseline model, DSAIN has increased the CTR by 2.70\%, the CPM by 2.62\%, and the GMV by 2.16\%. Currently, DSAIN has been deployed in the list-advertisement recommender system of the Meituan food delivery platform and is now serving hundreds of millions of users, which contributes to significant business revenue growth.

\section{Conclusion}
In this paper, we propose the concept of the situation and situational features for behaviors and then devise a novel CTR model DSAIN (Deep Situation-Aware Interaction Network). Given a sequence of user behaviors and the corresponding situational features, DSAIN reduces noise behaviors in the sequence, learns the high-quality embeddings for situational features by  taking into consideration multiple internal correlations, and aggregates them into the embedding of the behavior sequence for final CTR prediction. The results of offline experiments on three datasets and the online A/B test demonstrate the superiority of our DSAIN model.

%In this paper, we propose the concept of the situation and situational features for behaviors and then devise a novel CTR model DSAIN (Deep Situation-Aware Interaction Network). Given a user behavior and the corresponding situational features, DSAIN adopts four modules (i.e., the Behavior Denoising Module, the Situational Feature Encoder, the Correlation Fusion Module, and the Situation Aggregation Module) to learn the high-quality embeddings for situational features and then generate the embedding of the behavior sequence applied to the CTR prediction. The results of offline experiments on three datasets and the online A/B test demonstrate the superiority of our DSAIN model. 

% In the future, we will further investigate 

% there is a lot of room to explore how to efficiently and effectively integrate the side information into the interest representation space.

\begin{acks}
This research was supported by Meituan and the Natural Science Foundation of China under Grant No. 62072450.
\end{acks}

\bibliographystyle{ACM-Reference-Format}
\bibliography{sample-authordraft}


\end{document}